\documentclass[aps,twocolumn,showpacs]{revtex4}
\usepackage{amssymb}
\usepackage{graphicx}

\def\be{\begin{equation}}
\def\ee{\end{equation}}
\def\bea{\begin{eqnarray}}
\def\eea{\end{eqnarray}}

\def\H{{\cal H}}

\newcommand{\ket}[1]{\mbox{$| #1 \rangle$}}

\begin{document}

\title{Multipartite Bound Information exists and can be activated}
\author{A. Ac\'{\i}n$^{1}$, J. I. Cirac$^{2}$ and Ll. Masanes$^{3}$}
\affiliation{
$^{1}$Institut de Ci\`encies Fot\`oniques, Jordi
Girona 29, Edifici Nexus II, 08034 Barcelona, Spain\\
$^2$Max--Planck Institut f\"ur Quantenoptik, Hans--Kopfermann
Str. 1, D-85748 Garching, Germany\\
$^3$Dept. d'Estructura i Constituents de la Mat\`eria, 
Univ. de Barcelona, 08028 Barcelona, Spain}
\date{\today }

\begin{abstract}
We prove the conjectured existence of Bound
Information, a classical analog of bound entanglement,
in the multipartite scenario. We give examples 
of tripartite probability distributions from which it is impossible to extract any kind of secret key, even in the asymptotic 
regime, although they cannot be created by local 
operations and public communication. Moreover, we show 
that bound information can be activated: three honest parties 
can distill a common secret key from different
distributions having bound information. Our results
demonstrate that quantum information theory can provide useful
insight for solving open problems in classical information theory.
\end{abstract}

\maketitle

\bigskip

In 1993, Maurer introduced the following scenario for information-theoretically secure secret-key agreement \cite{Maurer}: several parties, including a possible adversary, share partially correlated (classical) information. The honest parties aim to establish a secret key, processing this information with local operations and public communication (LOPC). The secret key has to be completely uncorrelated to the adversary's information. 
Because information-theoretically secure secret bits cannot be created by LOPC, all the {\sl secrecy} has to come from the correlations that they initially have. Maurer's formulation shares many similarities with the standard scenario of entanglement manipulations in quantum information theory. There, several separated parties share many copies of a multipartite quantum state, which specifies the kind of quantum correlations existing among them and the environment. Their goal is to obtain pure-state entanglement applying only local operations and classical communication (LOCC). A pure state is uncorrelated to the environment. Then, the environment plays the same role as the adversary in cryptography. 
The analogy between both scenarios was first explored in Ref. \cite{GW} and later developed in Refs. \cite{CP,CMS}.

\bigskip

Given a state $\rho$ in a composite system of several parties, two fundamental questions in quantum information theory are: (i) can 
it be prepared by LOCC? and (ii) can pure-state entanglement be extracted from many copies of $\rho$ by LOCC? These questions, that still remain unsolved, define the separability and distillability problems (see for instance \cite{ent}). Despite the natural expectation that all entangled states were distillable, in 1998 the Horodecki family showed the existence of the so-called bound entangled states \cite{bound}. These are states from which it is impossible to extract pure-state entanglement although they cannot be created by LOCC. Following the analogy between the entanglement and key-agreement scenarios, Gisin and Wolf conjectured and gave evidence for the existence of a classical analog of bound entanglement, the so-called bound information \cite{GW}. This consists on information shared among several honest parties and an eavesdropper such that (i) it is impossible for the honest parties to extract a secret key, and (ii), this information cannot be distributed by LOPC. 

\bigskip

In this work we present the first provable examples of multipartite bound information. Remarkably, our examples can be {\sl activated} in the same sense as in the quantum case. That is, after LOPC processing different kinds of bound information, a secret key can be obtained. The intuition used to get these results entirely comes from already known examples of bound entangled states in three-qubit systems. Our work then, constitutes one of the first situations where the quantum information insight gives the answer to an open problem in classical information theory \cite{KW}. Indeed, up to now the flow of results has mainly been in the opposite direction, e.g. the quantum protocols for entanglement distillation of Ref. \cite{BDSW} were derived from existing classical protocols for key distillation. But before proving our results, let us review some known facts about secret-key distillation.

\bigskip

In his original formulation of the key-agreement problem, Maurer just considered the bipartite scenario: two honest parties (Alice and Bob) connected by an authentic but otherwise insecure classical communication channel, such that, a possible eavesdropper (Eve) learns the whole communication between them. Additionally, each party ---including Eve--- has access to correlated information given by repeated realizations of the random variables $A$, $B$ and $E$ (possessed by Alice, Bob and Eve respectively), jointly distributed according to $P(A,B,E)$. The goal for Alice and Bob is to obtain a common string of random bits for which Eve has virtually no information, i.e. a secret key. The maximal amount of secret key bits that can be asymptotically extracted per realization of $(A,B)$ used, is called the secret key rate, denoted by $S(A:B\parallel E)$. This quantity can be seen as the analog of the distillable entanglement, $E_d$ \cite{entan}. More recently, the so-called information of formation $I_{\mbox{\scriptsize form}}(A;B|E)$, has been introduced in \cite{RW} as the analog of the entanglement cost, $E_c$ \cite{entan}. Given $P(A,B,E)$, it can be understood as the minimal number of secret key bits asymptotically needed to generate each independent realization of $(A,B)$ ---distributed according to $P(A,B)$---, such that the information about $(A,B)$ contained in the messages exchanged through the public channel is at most equal to the information in $E$ \cite{inffdef}. A probability distribution can be established by LOPC if and only if $I_{\mbox{\scriptsize form}}=0$. Using these quantities, we can now define bound information. A probability distribution $P(A,B,E)$ contains bound information when the following two conditions hold \cite{notebound}:
\bea 
  S(A:B\parallel E)&=&0 \nonumber\\
  I_{\mbox{\scriptsize form}}(A:B|E)&>&0\ . 
\eea

A useful upper bound for $S(A:B\parallel E)$ is given by the so-called intrinsic information, introduced in \cite{MW}. This quantity, denoted by $I(A:B\downarrow E)$, will play a significant role in the proof of our results. The intrinsic information between $A$ and $B$ given $E$ is defined as:
\be
  I(A:B\downarrow E)\ =\ \min_{E\rightarrow \tilde{E}}\ I(A:B|\tilde{E})\ ,
  \label{intrinf} 
\ee
where the minimization runs over all possible stochastic maps $P(\tilde{E}|E)$ defining a new aleatory variable $\tilde{E}$. The quantity $I(A:B|E)$ is the mutual information between $A$ and $B$ conditioned on $E$. It can be written as
\be
  I(A:B|E)=H(A,E)+H(B,E)-H(A,B,E)-H(E) \ ,
\ee
where $H(X)$ is the Shannon entropy of the aleatory variable $X$. The intrinsic information also gives a lower bound for the information of formation \cite{RW}, thus
\begin{equation}
\label{ibounds}
  S(A:B\parallel E)\leq I(A:B\downarrow E)\leq I_{\mbox{\scriptsize form}}(A:B|E)\ .
\end{equation}

\bigskip

The generalization of Maurer's formulation to the multipartite scenario is straightforward. In our case, three honest parties ---Alice, Bob and Clare--- are connected by a broadcast public communication channel which is totally accessible to the eavesdropper ---Eve--- but that she cannot tamper. As it happens in entanglement theory, the generalization of the secret key rate ($E_d$) and the information of formation ($E_c$) to the multipartite case may not be univoque \cite{CMS}. Anyhow, the idea of multipartite bound information is unambiguous: a probability distribution $P(A,B,C,E)$ contains bound information if
\begin{enumerate}
\item No pair of honest parties ---even with the help of the third one--- can generate a secret key from many copies of $P(A,B,C,E)$. This also prevents the possibility of distilling a tripartite secret key \cite{tripartite}, because from it, a bipartite key between any pair of parties could be generated, giving a contradiction (see the note \cite{notekey}).

\item Its distribution by LOPC is not possible. More precisely, a large number of realizations of the aleatory variables $A$, $B$ and $C$ following the reduced probability distribution $P(A,B,C)$, cannot be distributed among Alice, Bob and Clare if the broadcasted messages are constrained to contain at most the information of the variable $E$ \cite{inffdef}.
\end{enumerate}
Having collected all these facts, let us prove the main result of this work, namely the existence of bound information. 

Our example of bound information is given by the following probability distribution, denoted by $P_1$:
\begin{center}\begin{large}
\begin{tabular}{|ccc|c|c|} \hline
  $\ A\ $ &$B$ &$\ C\ $ &$\ E\ $ &$P_1(A,B,C,E)$ \\ \hline
  0 &0 &0 &0 &$1/6$ \\
  0 &0 &1 &1 &$1/6$ \\
  0 &1 &0 &2 &$1/6$ \\
  1 &0 &1 &3 &$1/6$ \\
  1 &1 &0 &4 &$1/6$ \\
  1 &1 &1 &0 &$1/6$ \\
\hline
\end{tabular}
\vspace{10pt}
\end{large}\end{center}
This is the probability distribution that one obtains after measuring the three-qubit bound entangled state $\rho_1$, given in the Eq. (17) of Ref. \cite{DC}, in the computational basis. Note that $P_1(0,1,1)=P(1,0,0)=0$ and this distribution is invariant, up to a relabeling of $E$, under interchange of $B$ and $C$. In what follows, it is seen that from these correlations, it is impossible to extract a secret key between any pair of parties, even with the help of the third one. 

First, consider the bipartite splitting $AB-C$, where Alice and Bob are allowed to perform joint (secret) operations, i.e. they are connected by a private channel. It is easy to see that $I(AB:C|E)=p\,(E=0)\,I(AB:C|E=0)=1/3$. Now, applying the stochastic map $E\rightarrow \tilde{E}$ corresponding to: $1\rightarrow 0$, $4\rightarrow 0$ and identity for the rest of the values, we obtain $I(AB:C|\tilde{E})=0$. That is, the intrinsic information (\ref{intrinf}) vanishes, and because of (\ref{ibounds}) we have that
\begin{equation}
\label{zerosecr}
	S(AB:C\parallel E)=0\ .
\end{equation}
This implies that Clare cannot establish a secret key with Alice nor with Bob (even in the favorable situation where Alice and Bob are together). Because $P_1$ is symmetric with respect to $B$ and $C$, we also have that Bob cannot extract a key with Alice nor with Clare. Therefore, no secret key between any pair of parties can be generated from many copies of $P_1$ by LOPC.

\bigskip

Notice that $P_1$ contains some kind of secret correlations, although they are not distillable in the previous scenarios. This fact becomes manifest when we allow Bob and Clare to perform joint operations. In this case, we have again that $I(BC:A|E)=1/3$. But now, it is possible to construct a key distillation protocol achieving this rate: Bob and Clare announce publicly the cases where they have $B=C$, without saying the specific value. Each of these filtered realizations of $P_1$, that happen with probability $1/3$, contains one secret bit shared between $A$ and $BC$. Therefore, 
\begin{equation}
\label{secrform}
	S(A:BC\parallel E)=I(A:BC\downarrow E)=\frac{1}{3}\ .
\end{equation}
This condition cannot be satisfied by those probability distributions created by LOPC, since in this case $S=0$ for all the bipartite splitting of the honest parties. Hence, by definition, $P_1$ is an example of bound information, since it contains non-distillable secret correlations.

\bigskip

As we have seen, the secret correlations present in $P_1$ can be activated when a private channel is established between Bob and Clare. Indeed, the secret key given to these two parties allows to activate the already existing secret correlations with Alice. A similar phenomenon also happens in the quantum case, e.g. for the state $\rho_1$ that inspired the construction of $P_1$. An even more intriguing example of activation of bound entanglement consists of the fact that the tensor product, and even the mixture of bound entangled states can contain distillable entanglement \cite{DC,SST}. This process is sometimes called {\sl superactivation} of bound entanglement. In the next lines, we show the analog of superactivation for secret correlations. Again, our example is inspired by the results of Ref. \cite{DC}.

\bigskip

Consider the case in which the honest parties have access to 
a source of correlated information that supplies 
them with three probability distributions $P_1$, $P_2$ and $P_3$, 
where $P_2$ and $P_3$ are cyclic permutation of $P_1$, 
  \bea
  P_2(A,B,C,E) &=& P_1(B,C,A,E) \nonumber\\
  P_3(A,B,C,E) &=& P_1(C,A,B,E) .
  \eea
Of course, all these distributions contain bound
information. Using only LOPC, Alice, Bob and Clare can construct
an equally weighted mixture of $P_1$, $P_2$ and $P_3$:
  \be
  P_{\mbox{\scriptsize mix}} = \frac{1}{3} \left( P_1+P_2+P_3 \right)
  \ee
An equivalent scenario would consists of a source preparing randomly the three distributions, in such a way that the knowledge about the actual distribution is only accessible to Eve. The resulting distribution, $P_{\mbox{\scriptsize mix}}$, is detailed in the following table:
\begin{center}\begin{large}
\begin{tabular}{|ccc|c|c|} \hline
  $\ A\ $ &$B$ &$\ C\ $ &$\ E\ $ & $\ P_{\mbox{\scriptsize mix}}(A,B,C,E)\ $\\ \hline
  0 &0 &0 &0 &$1/6$ \\
  0 &0 &1 &1 &$1/9$ \\
  0 &1 &0 &2 &$1/9$ \\
  0 &1 &1 &3 &$1/9$ \\
  1 &0 &0 &4 &$1/9$ \\
  1 &0 &1 &5 &$1/9$ \\
  1 &1 &0 &6 &$1/9$ \\
  1 &1 &1 &0 &$1/6$ \\
\hline
\end{tabular}
\vspace{10pt}
\end{large}\end{center}
Actually, if one takes into account the total information accessible to the parties, Eve's symbol should be equal to $(E,i)$, where $i=\{1,2,3\}$ specifies the distribution $P_i$ and $E$ is associated to the triple of random variables $(A,B,C)$. However, it is easy to see that this distribution is equivalent to $P_{\mbox{\scriptsize mix}}$ from the point of view of Eve's information on Alice, Bob and Clare's symbols. 

Interestingly, $P_{\mbox{\scriptsize mix}}$ can be distilled into a tripartite key. To achieve this goal, the honest parties can use the repeated code protocol of Ref. \cite{Maurer}, generalized to the multipartite scenario. It consists of the following steps:
\begin{enumerate}
  \item Each party takes $N$ realizations of its own random
  variable:
    \bea
    && A_1, A_2, \ldots A_N\nonumber\\
    && B_1, B_2, \ldots B_N\nonumber\\
    && C_1, C_2, \ldots C_N ,
    \eea
  where $A_i,B_i,C_i$ are correlated according to $P_{\mbox{\scriptsize
  mix}}(A_i,B_i,C_i,E_i)$, for every value of $i$.
  \item Alice ---or any of the honest parties--- generates locally a
  random bit $s_A$, computes the numbers $X_i:=A_i+s_A$,
  where the sum is modulo 2, 
  for each value of $i$, and broadcasts through the public channel 
  the $N$-bit string:
    \be
    X_1, X_2, \ldots X_N .
    \ee
  \item Bob adds bitwise this string to his symbols $B_1,B_2, \ldots B_N$. If he obtains the 
  same value for all of them, $B_i+X_i=s_B ,\,\forall i$, 
  he accepts $s_B$ and communicates the acceptance to the other 
  parties. If not, the $N$ realizations of $P_{\mbox{\scriptsize
  mix}}$ are rejected. Clare does the same, accepting $s_C$ 
  only when $C_i+X_i=s_C ,\,\forall i$.
\end{enumerate}
For any accepted $N$-bit string only four cases
are possible: $A_i=B_i=C_i\ \ \forall i$, $\ A_i=B_i \neq
C_i\ \ \forall i$, $\ B_i=C_i \neq A_i\ \ \forall i$ or $\ C_i=A_i \neq
B_i\ \ \forall i$. The probability of being in the first case, once the string has been accepted by Bob and Clare, reads
  \be
  P(s_A=s_B=s_C|\mbox{\small accepted}) = \frac{\left( \frac{2}{6} \right)^N}
  {\left( \frac{2}{6} \right)^N+3\left( \frac{2}{9} \right)^N}\ ,
  \ee
which tends to one for large $N$. Thus, this protocol
allows the honest parties to correct all their errors since it 
only selects the 000 and 111 events.
Note that for these filtered events, Eve has $E=0$ whatever the
value of $(A,B,C)$ is. Therefore, she has no information about
$s_A$, so the parties end sharing a perfect secret bit \cite{notedist}. 
This proves that $P_{\mbox{\scriptsize mix}}$ is distillable, although
it has been generated by LOPC from three probability distributions 
that are non-distillable. We have then that bound information can be
activated with bound information. Let us also mention that this activation provides
{\sl per se} an alternative proof of the fact that the initial 
probability distribution, $P_1$, contains secret correlations.

\bigskip

To summarize, in this work we have proven the existence of 
bound information, a classical analog of bound entanglement
conjectured in \cite{GW}, in the tripartite scenario. 
The intuition for our proof 
comes from known examples of bound entangled 
states in three-qubit systems. We have also shown that bound
information, like bound entanglement, can be
activated: the probabilistic mixture of three 
distributions having bound information
gives a distillable distribution.
These results are straightforward generalizable to an arbitrary number of parties. Indeed, we have found several examples of probability distributions having bound secret correlations, which exhibit a wide variety of activation properties. These results will be given elsewhere.

\bigskip

Of course, it still remains as an open question whether
bound information exists in the bipartite scenario, i.e. 
to find probability distributions $P(A,B,E)$ such 
that $0=S<I_{\mbox{\scriptsize form}}$. The previous
evidence given in Ref. \cite{GW} is now significantly 
strengthened by our results. And if it exists, the next open 
problem would be to see whether bound information can be activated, 
as it seems to happen for bound entanglement in the 
bipartite scenario \cite{SST2}.

\bigskip

We would like to conclude mentioning the intriguing analogies that 
exist between privacy and entanglement. Very recently, it has been shown
that any entangling channel can be seen as a source of privacy \cite{AG} and that
a secret key can be extracted even from some non-distillable quantum states \cite{HO}. 
Our results indeed exploit this connection, and constitute one 
example of an almost unexplored application
of quantum information theory: the use of its formalism to solve 
open problems in classical information theory.

\bigskip
\bigskip

The authors are thankful to Nicolas Gisin, Renato Renner, Valerio Scarani 
and Stefan Wolf for sharing their insight with us.
This work has been supported by the EU (projects RESQ and QUPRODIS), 
the Kompetenzenznetzwerk ``Quanteninformationsverarbeitung", the ESF,
the grant 2002FI-00373 UB and the Generalitat de
Catalunya.


\begin{thebibliography}{99}
\bibitem{inffdef}
More precisely: given $P(A,B,E)$, if $C$ denotes the 
communication sent over the public channel for the preparation 
of $N$ events of $(A,B)$, there exists a channel for Eve mapping
$N$ realizations of $E$ into $C$, where $N\rightarrow\infty$.
\bibitem{notebound}
The initial definition of bound information given
in Ref. \cite{GW} consisted on a probability distribution
$P(A,B,E)$ such that $0=S(A;B||E)< I(A:B\downarrow E)$,
where $I(A:B\downarrow E)$ is defined in Eq. 
(\ref{intrinf}). However the two definitions are 
equivalent (R. Renner and S. Wolf, private
communication).
\bibitem{tripartite}
By a tripartite key we mean a common random bit possessed by each of the
three honest parties, about which Eve has virtually no information.
\bibitem{notekey}
This is analogous to the scenario of entanglement transformations. There: (i) three collaborating parties can transform a shared GHZ state, $(\ket{000}+\ket{111})/\sqrt 2$, into a singlet, $(\ket{00}+\ket{11})/\sqrt 2$, between any pair of parties, and (ii), if Alice shares singlets with Bob and Clare, she can locally prepare a GHZ state and teleport two of the three qubits to both of them. These two procedures have the following cryptographic analogs: (i) the sharing of a tripartite secret key allows the honest parties to obtain a bipartite key between any two of them. The third party should simply forget her information. And (ii), if a party ---say Alice--- shares one secret key with Bob and other with Clare, she can distribute any kind of secret correlations using one-time pad, e.g. a tripartite key. 
\bibitem{notedist}
Actually, the honest parties should run this protocol for a 
sufficiently large $N$, and then apply error correction and privacy amplification techniques for distilling the key with asymptotically finite rate.
\end{thebibliography}
\end{document}